%% file: lumi_paper_new.tex
\documentclass{elsart}

\usepackage{fancyhdr,graphicx}

\usepackage{rotating}

\usepackage{amssymb,amsmath}

\sloppypar
\usepackage{epsfig}

\begin{document}

\input{hvf_def}

\input{macros}


\newcommand{\bx}{\mbox{$BX$ }}

\newcommand{\mc}{\mbox{$MC$ }}

\newcommand{\xx}[1]{ }

\newcommand{\noxx}[1]{#1}

\newcommand{\lu}{\mbox{$\mathcal{L}$}}

\long\def\symbolfootnote[#1]#2{\begingroup%
\def\thefootnote{\fnsymbol{footnote}}\footnote[#1]{#2}\endgroup}

\begin{frontmatter}

\title{Luminosity determination at HERA-B}

\input{authors_epjc.tex}

\clearpage

\begin{abstract}

A detailed description of an original method used to measure the
luminosity accumulated by the HERA-B experiment for a data sample
taken during the 2002-2003 HERA running period is reported. We
show that, with this method, a total luminosity measurement can be
achieved with a typical precision, including overall systematic
uncertainties, at a level of  $5\%$ or better. We also report
evidence for the detection of $\delta$-rays generated in the target
and comment on the possible use of such delta rays to measure
luminosity.
\end{abstract}

\end{frontmatter}

\section{Introduction}
\label{sec:Introdu}
\symbolfootnote[0]{Corresponding Author: Marco.Bruschi@bo.infn.it}
\input{introduction.tex}

\section{The HERA accelerator and the target}

\label{sec:hera}

\input{hera.tex}

\section{The HERA-B detector and the data sample}

\label{sec:detector}

\input{detector.tex}

\section{The cross sections}

\label{sec:cross}

\input{cross.tex}

\section{General remarks on the luminosity determination}

\label{sec:General-remarks-luminos-determi}

\input{general_remarks.tex}

\section{The determination of $\lambda_{mb}$}

\label{sec:det_of_lambda}

\input{lambda.tex}

\section{Systematic uncertainties and checks}

\label{sec:Study-systema-uncerta}

\input{study_of_systematic.tex}

\input{dsect.tex}

\section{Summary and conclusions}

\label{sec:Conclus }

\input{conclusions.tex}

\section{Acknowledgments}

\label{sec:Acknowl}

\input{acknowledgments.tex}

\end{document}

%% file: hvf_def.tex
\newcommand{\B}{\ensuremath{{b}}}
\newcommand{\bjpsiX}{\ensuremath{ \B\to\jpsi X}}
\newcommand{\bjpsi}{\ensuremath{\B \to\jpsi}}
\newcommand{\bjpsill}{\ensuremath{\B \to\jpsi\to l^+l^-}}
\newcommand{\bjpsiee}{\ensuremath{\B \to\jpsi\to e^+e^-}}
\newcommand{\bjpsimm}{\ensuremath{\B \to\jpsi\to \mu^+\mu^-}}
\newcommand{\bbar}{\ensuremath{\B\overline{\B}}}
\newcommand{\sigbbar}{\ensuremath{\sigma(\bbar)}}
\newcommand{\pntobbar}{\ensuremath{pN\to \bbar }}
\newcommand{\ee}{\ensuremath{e^+e^-}}
\newcommand{\mm}{\ensuremath{\mu^+\mu^-}}
\newcommand{\bn}{\ensuremath{{\mathrm B}^0}}
\newcommand{\bnbar}{\ensuremath{\mathrm \overline{B}^0}}
\newcommand{\bnd}{\ensuremath{\mathrm B^0_d}}
\newcommand{\bndbar}{\ensuremath{\mathrm \overline{B}^0_d}}
\newcommand{\bns}{\ensuremath{\mathrm B^0_s}}
\newcommand{\bnsbar}{\ensuremath{\mathrm \overline{B}^0_s}}
\newcommand{\asym}{\ensuremath{frac{\Gamma_{B^0}(t)   - \Gamma_{\overline{B^0}}(t) }
                         {\Gamma_{B^0}(t)   + \Gamma_{\overline{B^0}}(t) } } }
\newcommand{\jpsi}{\ensuremath{\mathrm J/\psi}}
\newcommand{\jpsiX}{\ensuremath{\mathrm \jpsi X}}
\newcommand{\jpsill}{\ensuremath{ \jpsi\to l^+l^-}}
\newcommand{\jpsiee}{\ensuremath{ \jpsi\to e^+e^-}}
\newcommand{\jpsimm}{\ensuremath{ \jpsi\to \mu^+\mu^-}}
\newcommand{\ks}{\ensuremath{{\mathrm  K}^{\mathrm 0}_{\mathrm s}}}
\newcommand{\K}{\ensuremath{{\mathrm  K}^+}}
\newcommand{\xf}{\ensuremath{x_{\rm F}}}
\newcommand{\pt}{\ensuremath{p_{\rm T}}}

\newcommand{\lamb}{\ensuremath{\Lambda}}
\newcommand{\alamb}{\ensuremath{\overline{\Lambda}}}
\newcommand{\chic}{\ensuremath{\chi_C}}

\newcommand{\dm}{\ensuremath{\Delta m} }
\newcommand{\dmd}{\ensuremath{\Delta m_{\mathrm d}}}
\newcommand{\egev}{\ensuremath{\, \mathrm{GeV}}}
\newcommand{\mgev}{\ensuremath{\, \mathrm{GeV}/c^2}}
\newcommand{\pgev}{\ensuremath{\, \mathrm{GeV}/c}}
\newcommand{\emev}{\ensuremath{\, \mathrm{MeV}}}
\newcommand{\mmev}{\ensuremath{\, \mathrm{MeV}/c^2}}
\newcommand{\pmev}{\ensuremath{\, \mathrm{MeV}/c}}
\newcommand{\mum}{\ensuremath{\, \mu\mathrm m}}
\newcommand{\acp}{\ensuremath{a_{\mathrm CP}}}
\newcommand{\ccp}{\ensuremath{c_{\mathrm CP}}}
\newcommand{\ecp}{\ensuremath{\epsilon_{\mathrm B}} }

%% file: macros.tex
%
%
\newcommand{\ARNP}[3]      {Ann.\ Rev.\ Nuc.\ Part.\ Sci.~{\bf #1} (#2) #3}
\newcommand{\IEEE}[3]      {IEEE Trans.\ Nucl.\ Sci.~{\bf NS-#1} (#2) #3}
\newcommand{\NIM}[3]       {Nucl.\ Instr.\ Methods~{\bf A#1} (#2) #3}
\newcommand{\NPA}[3]       {Nucl.\ Phys.~{\bf A#1} (#2) #3}
\newcommand{\NPB}[3]       {Nucl.\ Phys.~{\bf B#1} (#2) #3}
\newcommand{\PLB}[3]       {Phys.\ Lett.~{\bf B#1} (#2) #3}
\newcommand{\PRC}[3]       {Phys.\ Rev.~{\bf C#1} (#2) #3}
\newcommand{\PRD}[3]       {Phys.\ Rev.~{\bf D#1} (#2) #3}
\newcommand{\PRL}[3]       {Phys.\ Rev.\ Lett.~{\bf #1} (#2) #3}
\newcommand{\SJNP}[3]      {Sov.~J.\ Nucl.\ Phys.~{\bf #1} (#2) #3}
\newcommand{\ZPC}[3]       {Z.~Phys.~{\bf C#1} (1#2) #3}
\newcommand{\EPJ}[3]       {Eur.\ Phys.\ J.~{\bf C#1} (#2) #3}
\newcommand{\CPC}[3]       {Comp.\ Phys.\ Comm.~{\bf #1} (#2) #3}
%
%
\newcommand{\hb} {\mbox{\sffamily HERA \protect\rule[.5ex]{1.ex}{.11ex} B}\ }
\newcommand{\hbp}{\mbox{\sffamily HERA \protect\rule[.5ex]{1.ex}{.11ex} B}}
\newcommand{\ra} {\mbox{$\mskip 3mu \rightarrow \mskip 5mu$}}
\newcommand{\mt} {\mbox{$p_\mathrm{T}$}}
\newcommand{\ptp}{\mbox{$\mt > 1.5~{\mathrm{GeV}/c}$}}
\newcommand{\mct}{\mbox{$M>4.5~{\mathrm{GeV}/c^2}$}}
\newcommand{\hpt}{\mbox{high-\mt}}
\newcommand{\nl} { \hfill\break }
\newcommand{\np} { \vfill\eject }
\newcommand{\vspitem}{\vspace{-2truemm}}

\newcommand{\al} {$\alpha$}
\newcommand{\be} {$\beta$}
\newcommand{\co} {$^{60}$Co}
\newcommand{\sr} {$^{90}$Sr}
\newcommand{\dc} {$^\circ$C}
\newcommand{\ga} {$\gamma$}
\newcommand{\dk} {$^\circ$K}
\newcommand{\ko} {k$\Omega\cdot$cm}
\newcommand{\flnu} {cm$^{-2}$}
\newcommand{\flxu} {cm$^{-2}$s$^{-1}$}
\newcommand{\neff} {$N_{eff}$}
\newcommand{\tflu} {$3\cdot 10^{14}$/cm$^2$}
\newcommand{\stflux} {$3\cdot 10^{7}$~cm$^{-2}$s$^{-1}$}

\newcommand{\eff}{\ensuremath{\varepsilon}}
\newcommand{\effB}{\ensuremath{\eff_B^{\jpsi}}}
\newcommand{\effBz}{\ensuremath{\eff_B^{\Delta z}}}
\newcommand{\effBt}{\ensuremath{\eff_B^{\rm tot}}}
\newcommand{\effP}{\ensuremath{\eff_P^{\jpsi}}}
\newcommand{\effPt}{\ensuremath{\eff_P^{\rm tot}}}
\newcommand{\effR}{\ensuremath{\eff_R}}
\newcommand{\bbjX}{\ensuremath{\bbar\ra\jpsiX}}
\newcommand{\jll}{\ensuremath{\jpsi\ra\dilepton}}
\newcommand{\lum}{\ensuremath{{\cal L}}}
\newcommand{\Br}[1]{\ensuremath{{\rm  Br}(#1)}}
\newcommand{\sigB}{\ensuremath{\sigma_B^A}}
\newcommand{\sigP}{\ensuremath{\sigma_P^A}}
\newcommand{\Dz}{\ensuremath{\Delta z}}
%
\newcommand{\bm}   {B~meson}                                            
\newcommand{\PB}   {\mbox{\ensuremath{\mathrm{B}}}}                     
\newcommand{\PBp}  {\mbox{\ensuremath{\mathrm{B}^+}}}                   
\newcommand{\PBm}  {\mbox{\ensuremath{\mathrm{B}^-}}}                   
\newcommand{\PMB}  {\mbox{\ensuremath{m_{\mathrm{B}}}}}                 
\newcommand{\PBz}  {\mbox{\ensuremath{\mathrm{B}^0}}}                   
\newcommand{\PaBz} {\mbox{\ensuremath{\overline{\mathrm{B}^0}}}}        
\newcommand{\PMBz} {\mbox{\ensuremath{m_{\mathrm{B}^0}}}}               
\newcommand{\PBzd} {\mbox{\ensuremath{\mathrm{B}^0_{\mathrm{d}}}}}      
\newcommand{\PMBzd}{\mbox{\ensuremath{m_{\mathrm{B}^0_{\mathrm{d}}}}}}  
\newcommand{\PaBzd}{\mbox{\ensuremath{\overline{\mathrm{B}^0_{\mathrm{d}}}}}} 
\newcommand{\PBzs} {\mbox{\ensuremath{\mathrm{B}^0_{\mathrm{s}}}}}      
\newcommand{\PMBzs}{\mbox{\ensuremath{m_{\mathrm{B}^0_{\mathrm{s}}}}}}  
\newcommand{\PaBzs}{\mbox{\ensuremath{\overline{\mathrm{B}^0_{\mathrm{s}}}}}} 

\newcommand{\PDz} {\mbox{\ensuremath{\mathrm{D^0}}}}                    
\newcommand{\PDps}{\mbox{\ensuremath{\mathrm{D^+_s}}}}                  
\newcommand{\PDms}{\mbox{\ensuremath{\mathrm{D^-_s}}}}                  

\newcommand{\PJgy}{\ensuremath{{\rm J}/\psi}}                                  
\newcommand{\PKp} {\mbox{\ensuremath{\mathrm{K}^+}}}                    
\newcommand{\PKm} {\mbox{\ensuremath{\mathrm{K}^-}}}                    
\newcommand{\PKpm}{\mbox{\ensuremath{\mathrm{K}^\pm}}}                  
\newcommand{\PKzS}{\mbox{\ensuremath{\mathrm{K^0_S}}}}                  
\newcommand{\PKzL}{\mbox{\ensuremath{\mathrm{K^0_L}}}}                  
\newcommand{\PKst}{\mbox{\ensuremath{\mathrm{K^*(892)}}}}               
\newcommand{\dilepton}{\mbox{\ensuremath{\ell^+ \ell^-}}}               
\newcommand{\epem}{\mbox{\ensuremath{\mathrm{e}^+ \mathrm{e}^-}}}       
\newcommand{\mpmm}{\mbox{\ensuremath{\mu^+ \mu^-}}}                     
\newcommand{\pppm}{\mbox{\ensuremath{\pi^+ \pi^-}}}                     
\newcommand{\jpsiks}{\mbox{\ensuremath{\PJgy \mskip 5mu \PKzS}}}        

\newcommand{\bjp}{\mbox{$\PBz \ra \jpsiks$}}                            
\newcommand{\bsbs}{\mbox{$\PBzs - \PaBzs$}}                             
\newcommand{\bpp}{\mbox{$\PBz \ra \pppm$}}                              
\newcommand{\bkp}{\mbox{$\PBz \ra \PKp \mskip 5mu \pi^-$}}              
\newcommand{\bpa}{\mbox{$\PBz \ra \pi^{\mp} \mskip 5mu a_1^{\pm}$}}     
\newcommand{\bdk}{\mbox{$\PBp \ra \PDz \mskip 5mu \PKp$}}               
\newcommand{\brk}{\mbox{$\PBp \ra \rho^0 \mskip 5mu \PKp$}}             
\newcommand{\bks}{\mbox{$\PBp \ra \pi^+ \mskip 5mu PKst$}}              
\newcommand{\brp}{\mbox{$\PBp \ra \pi^+ \mskip 5mu \rho^0$}}            
\newcommand{\bsk}{\mbox{$\PBzs \ra \mathrm{D^{\mp}_s} \mskip 5mu \PKpm$}}
\newcommand{\bsp}{\mbox{$\PBzs \ra \PDms \mskip 5mu \pi^+$}}            
\newcommand{\bsa}{\mbox{$\PBzs \ra \PDms \mskip 5mu 3\pi^{\pm}$}}       
%

%% file: authors_epjc.tex
I.~Abt$^{24}$,
M.~Adams$^{11}$,
M.~Agari$^{14}$,
H.~Albrecht$^{13}$,
A.~Aleksandrov$^{30}$,
V.~Amaral$^{9}$,
A.~Amorim$^{9}$,
S.~J.~Aplin$^{13}$,
V.~Aushev$^{17}$,
Y.~Bagaturia$^{13,37}$
% ...

\vspace{5mm}
\noindent
% ...
$^{9}${\it LIP Coimbra, P-3004-516 Coimbra,  Portugal~$^{f}$} \\
% ...
$^{11}${\it Institut f\"ur Physik, Universit\"at Dortmund, D-44221 Dortmund, Germany~$^{d}$} \\
% ...
$^{13}${\it DESY, D-22603 Hamburg, Germany} \\
$^{14}${\it Max-Planck-Institut f\"ur Kernphysik, D-69117 Heidelberg, Germany~$^{d}$} \\
% ...
$^{17}${\it Institute for Nuclear Research, Ukrainian Academy of Science, 03680 Kiev, Ukraine~$^{h}$} \\
% ...
$^{24}${\it Max-Planck-Institut f\"ur Physik, Werner-Heisenberg-Institut, D-80805 M\"unchen, Germany~$^{d}$} \\
% ...
$^{30}${\it DESY, D-15738 Zeuthen, Germany} \\
% ...
$^{37}${\it visitor from High Energy Physics Institute, 380086 Tbilisi, Georgia} \\
% ...
$^{d}$ supported by the Bundesministerium f\"ur Bildung und Forschung, FRG, under contract numbers 05-7BU35I, 05-7DO55P, 05-HB1HRA, 05-HB1KHA, 05-HB1PEA, 05-HB1PSA, 05-HB1VHA, 05-HB9HRA, 05-7HD15I, 05-7MP25I, 05-7SI75I \\
% ...
$^{f}$ supported by the Portuguese Funda\c c\~ao para a Ci\^encia e Tecnologia under the program POCTI \\
% ...
$^{h}$ supported by the National Academy of Science and the Ministry of Education and Science of Ukraine \\
% ...

%% file: introduction.tex
A precise determination of the luminosity
is required for the measurement of absolute cross sections.
The integrated luminosity ($\mathcal{L}$) is defined by
\begin{equation}\label{eq:lumgen}
 \mathcal{L}  =
\frac{{N_P }}{{\sigma_P }}
\end{equation}
where $N_P$ is the  number of events of a given process and
$\sigma_P$ is the corresponding cross section.
In the case of HERA-B, which is a forward spectrometer~\cite{HB_TDR,HBREP}
experiment, operated at the
920 GeV proton beam of the HERA accelerator at the DESY Laboratory
in Hamburg, the proton beam is bunched and interacts with a
nuclear target placed on the halo of the beam. The number of
proton-nucleus (pA) interactions per bunch crossing is subject to
statistical fluctuations. For HERA-B, as for all other
experiments having a bunched beam, the luminosity
can be expressed as:
\begin{equation}\label{eq:hb_lum}
\mathcal{L} = \frac{N_{BX}  \cdot \lambda }{\sigma }
\end{equation}
where $\lambda$ is the average number of interactions per
bunch crossing $BX$,
$N_{BX}$ is the number of beam bunches crossing the apparatus
and $\sigma$ is the interaction
cross section (for a more detailed discussion, see Sections
~\ref{sec:cross} and ~\ref{sec:General-remarks-luminos-determi}).
As a consequence, given the cross section of proton-nucleus
interactions, the luminosity can be measured by determining
$\lambda$ and $N_{BX}$. The average number of interactions per
$BX$ can be obtained from a fully unbiased sample of events in
various ways: by looking at inclusive quantities which are
proportional to the number of interactions in one event (such as
the number of tracks or the energy released in a calorimeter), by
counting the number of primary vertices or by counting the number
of empty events. The first method
has the advantage of entailing only a rather straightforward
analysis of the data,
but the signal corresponding to a single interaction must be
evaluated precisely and detector stability becomes a relatively
critical issue.
In the second method, the vertex reconstruction efficiencies
must be known precisely as well as the probability of erroneously
merging or splitting primary vertices during reconstruction. In
the third method, the distribution of the number of interactions per
bunch crossing must be either known or assumed and the efficiency
for detecting
non-empty events and the impact of noise events must be evaluated.\\
After careful studies the HERA-B Collaboration has decided to
exploit the method based on counting events with evidence of at
least one interaction (which is equivalent to the third method listed
above), since this method minimizes the systematic
error on the luminosity determination allowing to
achieve a final precision of about 5\%.\\
The paper is organized in the following way. In
Sections~\ref{sec:hera} and~\ref{sec:detector} the main features
of the HERA accelerator relevant for this analysis and the HERA-B
detector are briefly described. Section~\ref{sec:cross} summarizes
all of the published proton-nucleus cross section measurements
which are used for the luminosity determination. In
Section~\ref{sec:General-remarks-luminos-determi}
and~\ref{sec:det_of_lambda}, the relevant relations for the
determination of the luminosity are described. In
Section~\ref{sec:Study-systema-uncerta} we discuss the systematic
uncertainties and comment on delta ray production,
while in Section~\ref{sec:Conclus } we report the
results obtained for the interaction trigger (defined below) data
sample.

%% file: hera.tex
HERA is a double storage ring designed for colliding a 920 GeV proton beam with
a 26 GeV electron beam. Four interaction regions exist: two of them
house the general purpose {\it ep} detectors H1 and ZEUS, while the
other two accommodate the fixed target experiments HERA-B  and HERMES.
In the following we describe the beam parameters and the filling scheme used
during the HERA-B data taking period 2002-2003.

\begin{figure}[htb]
\begin{center}
\epsfig{file=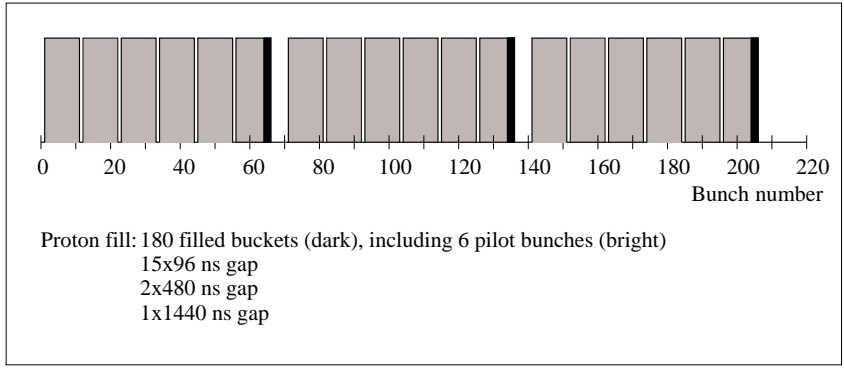,width=0.80\columnwidth}
\caption{\it Schematic representation of the bunch structure of a HERA
proton-ring fill.
    \label{fig:layout} }
\end{center}
\end{figure}

The typical proton current is 80 mA, distributed over
180 bunches with a typical bunch length of 1-2 ns. The proton bunches
are organized into 3x6 trains of 10 consecutive bunches each,
separated by one empty RF bucket. The detailed filling scheme is shown in
Fig.~\ref{fig:layout}.
In total there are 220 RF buckets with a spacing of 96 ns including
a gap of 15 empty buckets at the end to provide for a secure beam dump.
The average rate of filled bunch crossings is 8.52 MHz.

The target system~\cite{ehr00} consists of two stations of four wires
each. The wires are positioned above, below, and on either side of the
beam and are made from various materials including carbon, titanium
and tungsten. Both titanium and tungsten targets are wires with a diameter
of 50~$\mu$m, whereas the carbon target is a flat ribbon, 100~$\mu$m
perpendicular and 500~$\mu$m along the proton beam. 
The stations are separated by 40~mm along the beam
direction.  The wires are positioned individually in the halo of
the stored proton beam and the interaction rate for each inserted wire
is adjusted independently. Any number of wires can be operated simultaneously.
The luminosity measurement described herein applies exclusively to single
wire runs.

The steering of the target wires requires a fast and reliable system to
provide a counting rate proportional to the interaction rate up to the
highest interaction rates envisaged in the HERA-B design (40 MHz).
This is achieved by limiting the acceptance of the scintillation
counters used to detect interactions to $\sim 10^{-2}$.  
Stepping motors with a nominal step-size of 50 nm
controlled by a $10~$Hz steering loop provide a stable interaction rate.

One additional complication is that a fraction (typically a
few percent) of interactions not correlated
to any bunch~\cite{ehr01} was present. These interactions are due to
so-called coasting beam protons which have left the separatrix,
but are still circulating inside the machine, forming a component
of the beam halo. Based on test measurements,
the coasting beam can be regarded as a DC-current.
The fraction of coasting beam depends on the position of the
target and the history of the individual proton fill,
thus requiring an individual correction for each run.
As described in Section~\ref{sub:Backgro-estimat-other-related-systema-uncerta}, the relevant information can be derived from events
triggered by a pseudo-random generator.

%% file: detector.tex
\begin{figure*}[htb]
\centering
\resizebox{0.90\textwidth}{!}{%
\includegraphics{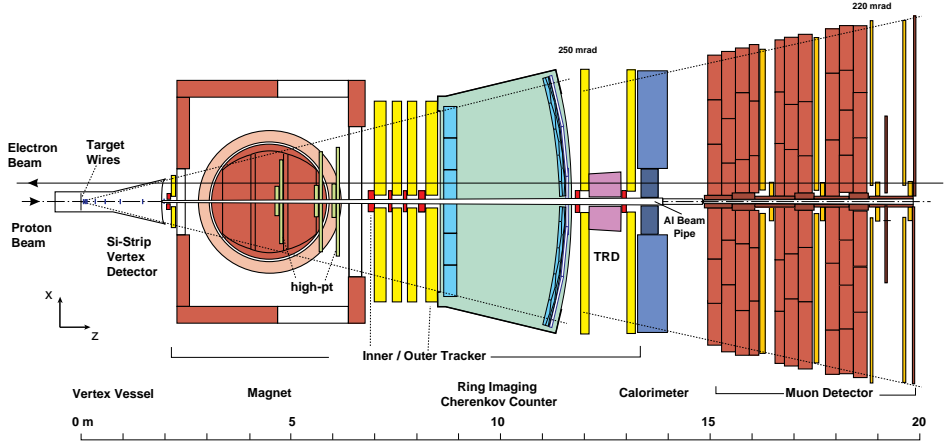}
} \caption{\it Top view of the HERA-B detector.}
\label{fig:1_detector}       
\end{figure*}

The HERA-B experiment is a forward magnetic spectrometer with an
acceptance extending from 15 to 220~mrad horizontally and to
160~mrad vertically.  This large angular coverage allows studies
in kinematic regions not accessible to previous fixed-target high
energy experiments. A top view of the detector is shown in Fig.
\ref{fig:1_detector}.
The first part of the spectrometer is devoted to tracking and
vertex measurements and consists of the target, a silicon vertex
detector, a magnet and a tracking system. The second part is
focused on particle identification and
includes a Ring Imaging Cherenkov detector, an electromagnetic
calorimeter and a muon
detector.\\
The vertex detector (VDS)~\cite{bau03} is placed between the target and the
magnet and divided in 8 stations. Each station consists of
four ``quadrants'' equipped with two double-sided silicon
microstrip detectors ($50\times70$~mm$^2$, 50 $\mu$m pitch) each.
This system provides a primary vertex resolution of
$\sigma_z \sim 500~\mu$m along the beam direction and
$\sigma_{x,y} \sim 50~\mu$m in the transverse plane.\\
A dipole magnet with a 2.13 Tm field-integral is positioned before
the main tracking system.  Each tracking station consists of several
planes of MSGC/GEM chambers placed near the beam pipe (Inner
Tracker, ITR)~\cite{zeu00} and several planes of Honeycomb Drift
chambers which cover the rest of the acceptance (Outer Tracker,
OTR)~\cite{alb05}. The detector segmentation is designed to cope
with the particle flux variation with the distance from the beam
pipe. Typical
momentum resolutions of $\delta_p /p \sim 1\%$ are achieved.\\
The particle identification of charged tracks (protons, kaons,
etc.) is provided by a Cherenkov detector (RICH) installed
downstream of the magnet. A $\beta \sim 1$ particle traversing the
RICH detector produces an average of about 33 hits~\cite{ari04}.
The electromagnetic calorimeter (ECAL)~\cite{avo01}, which
provides electron pretrigger seeds and $e/\pi$ separation, is
installed after the RICH and the tracking system. The ECAL is a
Shashlik sampling calorimeter with Pb or W as absorber and
scintillator as active material. In order to follow the steep
radial dependence of the particle density, the calorimeter has
been structured in three sections (Inner, Middle and Outer) with
differing granularities. The Muon detector (MUON)~\cite{eig01}
provides the muon pretrigger seeds and the muon identification, and
is located in the most downstream part of the detector. It
consists of four superlayers embedded in an iron loaded concrete
absorber. The sensitive area close to the beam pipe is covered by
pixel chambers, while in the rest of the
acceptance, tube chambers are used.\\
The flexibility of the trigger system~\cite{bala02} allows the
implementation of a large variety of trigger configurations. The
methods described in this paper have been used to determine the
integrated luminosity of the interaction trigger (IA) data sample.
The IA trigger selects events with at least one inelastic
interaction in the target, by requiring either that the RICH has
more than a minimum number of hits (20) or that the ECAL has more
than a minimum (1 GeV) energy deposition. The total collected
statistics is about 220 million events, with an average data
acquisition (DAQ) rate larger than 1000 Hz. During the data
acquisition a sample of randomly trigger events (Zero-Bias) was
acquired in parallel to the IA trigger at a rate of few Hz,
allowing the possibility to check the trigger acceptance and
stability. Moreover the Zero-Bias event sample has been
extensively used in the luminosity determination. Due to the fact
that the same data stream was used both for the various physics
analyses and for the determination of the recorded luminosity,
the dead time of the DAQ system cancels exactly and can thus be ignored.

%% file: cross.tex
The total $pA$ cross section $\sigma_{tot}$ can be divided into
elastic ($\sigma_{el}$) and inelastic ($\sigma_{inel}$)
contributions:
\begin{equation}
\sigma_{tot}=\sigma_{el}+\sigma_{inel}=\sigma_{el}+\sigma_{mb}+\sigma_{tsd}+\sigma_{bsd}+\sigma_{dd} \; .
\label{eq:total}
\end{equation}
In this context, the cross section $\sigma_{el}$ is regarded as the sum
of the elastic ($pA\rightarrow pA$) and quasielastic contribution
($pA\rightarrow pA^*$).
The inelastic cross section includes a minimum bias part ($mb$) and a
diffractive part which can be further subdivided into target single
diffractive ($tsd$, $pA\rightarrow pY$), beam single diffractive
($bsd$, $pA\rightarrow XA$) and
double diffractive ($dd$, $pA\rightarrow XY$) contributions.

The values for the total and inelastic cross sections
reported in Table~\ref{table:joaoboris} were obtained using the method
of \cite{joao} with one exception: in order to minimize a possible
systematic bias, we do not apply the $A^\alpha$ scaling law adopted
there. Instead we use the experimental results on
carbon and tungsten nuclei for the total cross section of~\cite{murthy}
and for the inelastic cross sections
of~\cite{carroll,fumuro,roberts}.
All of these measurements were obtained at beam momenta ranging from 180 to 400
GeV/c and have to be scaled to 920 GeV/c with the prescription
given in \cite{pdg}. Due to the absence of data on titanium,
the $A^\alpha$ scaling law is applied only
to interpolate the Al and Fe data of the experiments quoted.
The elastic cross sections are obtained using Equation~\ref{eq:total}.
The single diffractive cross sections are taken from~\cite{joao}.
The experimental results can be compared to a theoretical calculation performed
in the framework of the Glauber-Gribov theory~\cite{boris}. Both total
and inelastic cross sections agree well within 5\%, while the
diffractive contributions exhibit larger discrepancies.
As suggested by~\cite{boris2}, an average of Model III and Model IV
of~\cite{boris} is used for this comparison. 
\footnote{Both models are based on the saturated form of the dipole
cross section and provide a more realistic description compared to
Models I and II which assume a quadratic dependence.}
The double diffractive part, being neglected in~\cite{joao}, is taken
from~\cite{boris}. The quoted errors cover the difference between both models.
The minimum bias cross section is derived by subtracting all diffractive
contributions from the inelastic cross section with an error given by
the quadratic sum of the component errors.

\begin{table}[ht]
\begin{center}
\begin{tabular}{|l|c|c|c|}
\hline \hline cross section (mb)&  C  &  Ti  &  W
\\ \hline \hline
$\sigma_{tot}$  & 351.6 $\pm$ 4.0 & 1045.$\pm$ 30. & 2913. $\pm$ 43.
\\ \hline
$\sigma_{inel}$ & 250.7 $\pm$ 2.6 & 682.5$\pm$ 5.6 & 1788. $\pm$ 23.
\\ \hline
$\sigma_{el}  $ & 100.9 $\pm$ 4.8 & 362.5$\pm$ 30.5 & 1125. $\pm$ 49.
\\ \hline
$\sigma_{bsd} $ &   8.4 $\pm$ 1.7 &  21.7$\pm$ 4.3 &  41.2 $\pm$ 8.2
\\ \hline
$\sigma_{tsd} $ &   9.2 $\pm$ 2.3 &  14.9$\pm$ 3.8 &  23.9 $\pm$ 6.3
\\ \hline
$\sigma_{dd}  $ &   0.7 $\pm$ 0.3 &   1.1$\pm$ 0.5 &   1.5 $\pm$ 0.6
\\ \hline \hline
$\sigma_{mb}$   & 232.4 $\pm$ 3.9 & 644.8$\pm$ 8.1 & 1721. $\pm$ 26.
\\ \hline \hline
\end{tabular}
\vspace*{0.3cm}
\caption{\it Summary of the proton nucleus cross sections for C, Ti
and W nuclei. Details are given in the text.
\label{table:joaoboris}}
\end{center}
\end{table}

The total trigger efficiency can be expressed as
\begin{equation}
\label{eq:eff} \varepsilon_{tot}  = \frac{{ \varepsilon_{el} \cdot
\sigma_{el} + \varepsilon_{mb} \cdot \sigma_{mb} +
\varepsilon_{bsd} \cdot \sigma_{bsd} + \varepsilon_{tsd} \cdot
\sigma_{tsd} + \varepsilon_{dd} \cdot
\sigma_{dd}}}{{\sigma_{tot}}}  \; ,
\end{equation}
where $\varepsilon_{x}$ is the efficiency for triggering on process $x$. The
trigger efficiencies are determined from Monte Carlo simulation.
FRITIOF 7.02~\cite{fritiof} is used to generate minimum bias events in pA
interactions, while diffractive events are generated by PYTHIA 5.7~\cite{pythia}
which, however, has the disadvantage of not taking
into account nuclear effects. Nonetheless, since diffractive cross sections
are small compared to the minimum bias cross section, they contribute little
to the recorded sample and nuclear effects for diffractive events can
be safely neglected.

The detector response is simulated by the GEANT 3.21
package~\cite{GEANT}.  Realistic detector efficiencies, readout noise
and dead channels are taken into account. The simulated events are
processed by the same reconstruction codes as the data. The resulting
interaction trigger efficiencies are summarized in
Table~\ref{table:pythiaall}. It has been checked that the results for
the {\it pp} processes do not depend on the wire position. The small
increase of the minimum bias efficiency with increasing atomic mass number
is correlated to the increasing track multiplicity.

\begin{table}[ht]
\begin{center}
\begin{tabular}{|l|c|c|c|c|c|}
\hline \hline
Process & Generator & pp  & C  &  Ti  &  W
\\ \hline \hline
$\varepsilon_{el}  $ & PYTHIA & 0.003  & & &
\\ \hline
$\varepsilon_{bsd} $ & PYTHIA & 0.583  & & &
\\ \hline
$\varepsilon_{tsd} $ & PYTHIA & 0.370  & & &
\\ \hline
$\varepsilon_{dd}  $ & PYTHIA & 0.578  & & &
\\ \hline
$\varepsilon_{mb}  $ & PYTHIA & 0.941  & & &
\\ \hline
$\varepsilon_{mb}  $ & FRITIOF &           & 0.933  & 0.953  & 0.970
\\ \hline
$\varepsilon_{tot} $ & Eq.~\ref{eq:eff}& & 0.642  & 0.607  & 0.586
\\ \hline
$K_A$ & Eq.~\ref{eq:kappa} &             & 0.960  & 0.969  & 0.978
\\ \hline \hline
\end{tabular}
\vspace*{0.3cm}
\caption{\it IA trigger efficiencies for the various processes.
$K_A$ is defined in Equation~\ref{eq:kappa}.
\label{table:pythiaall}}
\end{center}
\end{table}

From these numbers we can conclude that the elastic contribution is negligible
and that the diffractive processes are suppressed. The dominance of the
minimum bias part can be illustrated by calculating its detectable fraction
$K_A$
\begin{equation}
\label{eq:kappa}
K_A  = \frac{\sigma_{mb} \cdot \varepsilon_{mb} }
{\sigma_{tot} \cdot \varepsilon_{tot} }
\end{equation}
given in Table~\ref{table:pythiaall}. The impact of the uncertainties
on luminosities and trigger efficiencies will be discussed in
Section~\ref{evaluation}.

%% file: general_remarks.tex
In the following, the luminosity given by Equation~\ref{eq:hb_lum}
will be expressed in terms of the total number of events
satisfying the IA trigger ($N_{IA}$), the average number of
interactions per bunch crossing ($\lambda_{tot}$), the trigger
efficiency per single interaction ($\epsilon_{tot}$) and the total
hadronic cross section ($\sigma_{tot}$). In order to do this, two
assumptions are made:
\begin{itemize}
\item the number of interactions per filled bunch can be
described by a single Poisson distribution $P(n,\lambda_{tot})$,
for all bunch crossings in a given data run:
\begin{equation}\label{eq:poisson}
P(n,\lambda_{tot} ) = \frac{{\lambda ^n _{tot} e^{ - \lambda_{tot}
} }}{n!},
\end{equation}
\item
the trigger efficiency for $n$ interactions, $(\varepsilon_{tot})_n$,
is given by:
\begin{equation}\label{eq:hypeffic}
\left( {\varepsilon _{tot} } \right)_n  = 1 - (1 - \varepsilon
_{tot} )^n
\end{equation}
where $\varepsilon_{tot}$ is the trigger efficiency for single
interaction.
\end{itemize}
A test of the validity of the first assumption is discussed in
Section~\ref{Poisson_test}, while the second assumption
has been checked in Monte Carlo studies and by checking the
dependence of the measured $\lambda_{tot}$ with interaction rate
as measured by the target steering scintillator hodoscopes (see
Section~\ref{lambda_mb_vs_rate}).

With these assumptions, the total number of recorded triggers
resulting from interactions in the target, $N_{IA}$, is given by:
\begin{equation}\label{eq:nia}
N_{IA}  = N_{BX}  \cdot \sum\limits_{n = 0}^\infty  {\left(
{P(n,\lambda _{tot} ) \cdot \left( {\varepsilon _{tot} } \right)_n }
\right)}  = N_{BX}  \cdot \left( {1 - e^{ - \varepsilon _{tot} \cdot
\lambda _{tot} } } \right)
\end{equation}
where $N_{BX}$ is the total number of $BX$s considered. From this
equation, given the general relationship of
Equation~\ref{eq:hb_lum}, we finally obtain
\begin{equation}
{\mathcal{L}_{tot}}=\frac{N_{IA}\cdot\lambda_{tot}}
{(1-e^{-\varepsilon_{tot}\cdot\lambda_{tot}})\cdot\sigma_{tot}} .
\label{eq:lumi}
\end{equation}
Because the product $\varepsilon \cdot\lambda$ is typically
$\approx 10\%$ for our data taking conditions, the measured
luminosity is, to first order, inversely proportional to the
trigger efficiency and the cross section, while the average number
of interactions, $\lambda_{tot}$, enters only as a second order
correction.

In Equation~\ref{eq:lumi}, $N_{IA}$ can be expressed as a function
of the number of recorded triggers ($N_{tape}$) and of the number
of background events ($N_{bkg}$):
\begin{equation}\label{eq:fthresh}
N_{IA}  = N_{tape} - N_{bkg}  = N_{tape}  \cdot \left( {1 - f_{bkg}} \right)
\end{equation}
where $f_{bkg}$ is the fraction of background events in the sample
(see Section~\ref{sub:Backgro-estimat-other-related-systema-uncerta}).

Since, as discussed in Section~\ref{sec:cross}, the recorded event sample
is dominated by minimum bias interactions, the luminosity can be expressed as a
function of minimum bias quantities
\begin{equation}\label{eq:luminew}
\mathcal{L}_{tot} = \frac{N_{tape}\cdot \left({1 - f_{bkg}}
\right) \cdot \lambda _{mb} } {\left( {1 - e^{ - \varepsilon _{mb}
\lambda_{mb} } } \right) \cdot \sigma _{mb}} \cdot K_A ,
\end{equation}
where $K_A$ is defined in Section~\ref{sec:cross} and
$\lambda_{mb}$ is defined in Section~\ref{sec:det_of_lambda}. This
is the final expression which will be used to determine the
luminosity for each run.

%% file: lambda.tex
The determination of  $\lambda_{mb}$ relies on the pseudo random
trigger data sample acquired in parallel to the $IA$ trigger and
the  large-acceptance detectors which constitute the spectrometer.
Specifically, $\lambda_{mb}$ is obtained by combining the
information from a variety of  subdetectors to also provide a
cross-check of the stability of the result and the systematic
uncertainties due to the detector response and the event model of
the Monte Carlo. Only the filled bunches of HERA were considered.

The average number of $IA$ per $BX$ can be evaluated with respect
to any subdetector observable $X$ which depends linearly on the
interaction rate, by exploiting the following definition:
\begin{equation}\label{eq:lambda_new}
\lambda_{mb}  =  - \frac{1}{{\varepsilon_{mb} (X)}} \cdot \ln
\left( {1 - \frac{{N_X}}{{N_{BX}}}} \right)
\end{equation}
where $N_X$ is the number of events with observable $X$ above a
certain threshold  and $\varepsilon (X)$ is the corresponding
efficiency (i.e. probability that an interaction will result in
$X$ being above threshold) as evaluated from the {\sc FRITIOF}
simulation. 
To avoid possible confusion, we note that $\lambda_{mb}$
as defined by Equation~\ref{eq:lambda_new} is close to but not
equal to the average number of minimum bias interactions per $BX$. 
With this definition, Equation~\ref{eq:luminew} is nonetheless exact.
The
sample of random trigger events for all runs is sufficiently large
that the statistical error is always negligible compared to the
systematic error estimate.
The list of the observables $X$ used for the determination of
$\lambda_{mb}$ is given in Table~\ref{table:description}. As can
be seen, two subdetectors are directly involved in this method,
namely RICH and ECAL, while the VDS and the OTR are indirectly
involved when the number of reconstructed tracks is considered.

\begin{table}[h]
\begin{center}
\begin{tabular}{|l|l|}
\hline \hline
    X & DESCRIPTION \\
\hline \hline
hrich & number of reconstructed hits in the RICH detector \\
\hline
e(ECAL,inner) & total energy deposition in the ECAL inner section (GeV) \\
\hline
e(ECAL,middle) & total energy deposition in the ECAL middle section (GeV) \\
\hline
e(ECAL,outer) & total energy deposition in the ECAL outer section (GeV) \\
\hline
e(ECAL,ECAL) & total energy deposition in full ECAL (GeV) \\
\hline
nclus(ECAL) & number of reconstructed electromagnetic clusters in ECAL \\
\hline
hecal & number of hit towers in ECAL \\
\hline
 ntra & number of reconstructed tracks (VDS+OTR) \\
\hline
\hline
\end{tabular}
\vspace*{0.3cm}
\caption{\it Description of the eight $X$ quantities used to
determine $\lambda_{mb}$.
\label{table:description}}
\end{center}
\end{table}
It is important to note that for the determination of
$\lambda_{mb}$, no reconstructed quantity associated only with the
$VDS$ is used. The reason for this
 is that all such quantities were found to be
 sensitive to the presence of $\delta$-rays generated by
 the proton beam in the target, as will be further
 discussed in Section~\ref{sec:delta}.

\begin{figure}[h]
\centering
\includegraphics*[bb=2 270 520 520,width=\textwidth,clip=1]{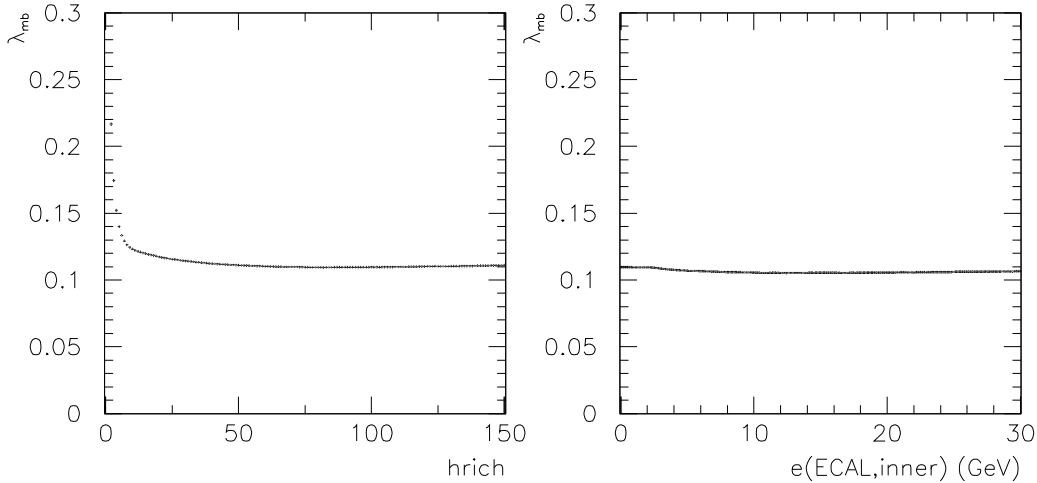}
\caption{\it Two typical distributions for
the measured average number of interactions per bunch crossing as
a function of the cut on different reconstructed quantities for
the tungsten target wire. The distributions for $RICH$
(number of reconstructed hits per event , -$hrich$-), $ECAL$
(total energy released per event in the inner section of the
calorimeter, -$e(ECAL,inner)$-) are shown.}
\label{fig:xdist}
\end{figure}
The $\lambda$ values calculated with  Equation~\ref{eq:lambda_new}
as a function of  two of the $X$ observables (namely $hrich$ and
$e(ECAL,inner)$) defined in Table~\ref{table:description} are
shown in Fig.~\ref{fig:xdist}. These plots show a common feature:
the existence of a broad stationary point for $\lambda_{mb}$. The
left plot in Figure~\ref{fig:xdist} shows the $\lambda_{mb}$
values obtained as a function of the threshold applied on the
number of hits ($hrich$) seen in the $RICH$ detector. The rise at
small values of $RICH$ hits is due to noise in the detector, while
the smooth increase for large number of $RICH$ hits is mainly due
to the fact that the Monte Carlo does not precisely reproduce the
$RICH$ hit multiplicity per event, although the resulting value of
$\lambda_{mb}$ is nearly independent of the threshold over a wide
range. This trend is confirmed, in a more or less pronounced way
(see e.g. the right plot of the same Figure), also for the other
variables listed in Table~\ref{table:description}.\\
 As a
consequence, for each observable, as the best estimate of
$\lambda_{mb}$ the value ($\lambda_{min}(X)$) is taken to be its
minimum value. The best evaluation of $\lambda_{mb}$ is then
defined as:
\begin{equation}\label{eq:lambda'exp}
\lambda_{mb}  = \sum\limits_{X = 1,8} {\frac{{\lambda
_{min}(X)}}{8}}
\end{equation}
The values of $\lambda _{min} ( X )$ obtained from  all are in
good agreement and their spread is used as a measure of the
systematic uncertainty.

%% file: study_of_systematic.tex
\subsection{General Considerations}
\label{sub:general_considerations}
According to Equation~\ref{eq:luminew} and the assumption on which
Equation~\ref{eq:poisson} is based, the following systematic
uncertainties must be taken into account:
\begin{itemize}
\item the uncertainty on $K_A$, arising from the Monte Carlo (\mc)
  event model and the poorly known observation probability of
  diffractive processes;
\item the uncertainty on the determination of ${\lambda _{mb}}$;
\item the uncertainty associated with deviations of the interaction
  probability distribution from the assumed Poisson distribution (e.g.
  due to the uneven filling of bunches);
\item the uncertainty associated with possible out-of-time
  interactions (i.e. coasting beam interactions) and fake triggers
  from detector noise.
\end{itemize}
No systematic uncertainty due to reconstruction efficiency appears
on this list since possible systematic biases due to
reconstruction are included in the systematic uncertainty assigned
to $\lambda_{mb}$.

The determination of $\mathcal{L}_{tot}$ could, however, be biased if
the on-line trigger does not operate according to expectations.
Possible triggering errors are checked for by using special bits
written into the event record to indicate the trigger decision. The
online trigger requirements are imposed offline on the sample of
random trigger events taken with each $IA$ trigger run and compared to
the the online decision. It is found that there is no significant
inefficiency from the online trigger while the percentage of spurious
triggers due to electronic misbehavior is typically at level of a few
per thousand. Some additional masking of noisy channels is found to be
necessary, although the effect on the trigger acceptance is
negligible.  We conclude that the trigger performed according to
expectations and introduces no additional biases on the measurement of
$\mathcal{L}_{tot}$.

The general expression for the squared relative uncertainty on the
luminosity follows from Equation~\ref{eq:luminew}:
\begin{equation}\label{eq:systema}
\left( {\frac{{\delta \mathcal{L}_{tot} }}{{\mathcal{L}_{tot} }}}
\right)^2 = \frac{{\delta K_A }}{{K_A }} \oplus \frac{{\delta
\lambda _{mb} }}{{\lambda _{mb} }} \oplus  \left( {\frac{{\delta
\mathcal{L}}}{\mathcal{L}}} \right)_{bkg} \oplus \left(
{\frac{{\delta \mathcal{L}}}{\mathcal{L}}} \right)_{Pois}\,,
\end{equation}
which is the quadratic sum of the  relative systematic
uncertainties on $K_A$ (see Section~\ref{evaluation}), on
$\lambda_{mb}$ (see Section~\ref{sub:Detector Efficiency}), on the
background and on the Poisson assumption (see
Equation~\ref{eq:poisson}). The last two sources of systematic
uncertainty will be discussed in Sections ~\ref{Poisson_test} and
~\ref{sub:Backgro-estimat-other-related-systema-uncerta}
respectively. Finally, to separate the measurement uncertainties
from the uncertainties on the present knowledge of the total cross
section ($K_A$ term), the following quantity (which will be used
in Section~\ref{sec:Conclus }) is defined:
\begin{equation}\label{eq:systdet}
\left( {\frac{{\delta \mathcal{L}_{tot} }}{{\mathcal{L}_{tot} }}}
\right)_{det}^2 = 
{\frac{{\delta \lambda _{mb} }}{{\lambda _{mb}}}} 
\oplus 
\left( {\frac{{\delta \mathcal{L}}}{\mathcal{L}}} \right)_{bkg}
\oplus 
\left( {\frac{{\delta \mathcal{L}}}{\mathcal{L}}}\right)_{Pois}\,.
\end{equation}

\subsection{Uncertainty on $K_A$}
\label{evaluation}

The relative uncertainty $(\frac{\delta{K_A}}{K_A})$ on the detectable
fraction $K_A$ (defined in Equation~\ref{eq:kappa}) depends on the
uncertainties of the cross sections, quoted in
Table~\ref{table:joaoboris}, and on the trigger efficiencies of the
various production processes. Given the poor knowledge of the
structure of final states produced by diffractive processes, the
trigger efficiencies are assumed to be fully unknown but limited to
the range from 0 to 1. Thus, an error of
\mbox{$\sigma_{\varepsilon_{bsd}} = \sigma_{\varepsilon_{tsd}} =
  \sigma_{\varepsilon _{dd} } =\frac{1}{{\sqrt {12} }}$} is assigned.
The error on $\varepsilon _{mb}$ is not included here as discussed in
Section~\ref{sub:general_considerations}. The resulting uncertainties
are summarized in Table~\ref{table:correction_factor}.
\begin{table}[h]
\begin{center}
\begin{tabular}{|c|c|c|}
\hline \hline
      &    $K_A$ & $\frac{\delta{K_A}}{K_A}$  \\
\hline \hline
    C & 0.960 & 0.023 \\
\hline
   Ti & 0.969 & 0.018 \\
\hline
    W & 0.978 & 0.016 \\
\hline \hline
\end{tabular}
\vspace*{0.3cm}
\caption{\it Detectable fractions and their relative uncertainty for
carbon, titanium and tungsten.
\label{table:correction_factor}}
\end{center}
\end{table}

\subsection{Uncertainty on $\lambda_{mb}$ }
\label{sub:Detector Efficiency}

The method used to determine $\lambda_{mb}$ is influenced by the
Monte Carlo description of the HERA-B detector as well as the event
model of the event generator. The resulting uncertainty on
$\lambda_{mb}$ is taken to be the $rms$ spread of the
$\lambda_{mb}$ values calculated with
Equation~\ref{eq:lambda'exp}. The typical values obtained are
\begin{equation}\label{eq:systonlam}
\frac{\delta_{\lambda_{mb}}}{\lambda_{mb}}\simeq 0.04\,,
\end{equation}
or better, depending on the target material.\\
Possible sources of systematic uncertainty related to the way
$\lambda_{mb}$ is determined have been investigated. For example,
examination of the observables given in
Table~\ref{table:description} shows that six of them involve
ECAL and are thus possibly subject to correlated systematic
effects while RICH and VDS+OTR appear only with one variable
each. For this reason an alternative quantity $\lambda'_{mb}$ is
defined as:
\begin{equation}\label{eq:lambda''exp}
\lambda ^{'} _{mb }  = \frac{{\lambda _{opt} \left( 1 \right) +
\sum\limits_{X = 2,7} {\frac{{\lambda _{opt} \left( X \right)}}{6}
+ \lambda _{opt} \left( {8} \right)} }}{3}.
\end{equation}
The relative discrepancy of $\lambda_{mb}$ and $\lambda'_{mb}$ is
given by:

\begin{equation}\label{eq:deltalambda}
\left(\frac{\delta \lambda_{mb}}{\lambda_{mb}}\right)_{method}  =
\frac{\lambda _{mb }  - \lambda{'} _{mb }}{\lambda _{mb }}\,.
\end{equation}
 The mean value of the  distribution of this quantity is found to
 be statistically compatible with zero and its $rms$ width is $\approx 0.006$
or better, depending on target material. The smallness of this
term, compared to the overall systematic uncertainty on
$\lambda_{mb}$ (see Equation~\ref{eq:systonlam}), shows that the
calculation of $\lambda_{mb}$ is insensitive to the relative
weights given to the various methods.

\subsection{Uncertainty on the  distribution of the number of $IA$ per $BX$}
\label{Poisson_test}

One important
assumption is that the number of interactions per HERA machine
bunch follows a Poisson distribution (see
Equation~\ref{eq:poisson}). In order to evaluate the systematic
uncertainty associated to the non-Poisson behavior of the $BX$
population, the total luminosity can alternatively be measured
for each run, also as a sum over all the $BX$ contributions, i.e.:

\begin{equation}\label{eq:ltotbx}
\left( {\mathcal{L}_{tot} } \right)_{BX}  = \frac{K_A}{{\sigma
_{mb}}} \cdot \sum\limits_{i = 1}^{180} {\frac{{N_{IA_i } \cdot
\lambda_{mb_i} }}{{\left( {1 - e^{ -
\varepsilon_{mb}\lambda_{mb_i}} } \right)}}}\,.
\end{equation}

The quantity $\left( {\mathcal{L}_{tot} } \right)_{BX}$ is then
compared with the total luminosity calculated according to
the basic procedure
(see Equation~\ref{eq:luminew}). In this way, we can define the
systematic uncertainty due to the non Poisson behavior of the beam
as:

\begin{equation}\label{eq:deltaltot/ltotpois}
\left( {\frac{{\delta \mathcal{L} }}{{\mathcal{L} }}}
\right)_{Pois} = \frac{{{\mathcal{L}_{tot} } - \left(
{\mathcal{L}_{tot} } \right)_{BX} }}{{{\mathcal{L}_{tot} } }}\,.
\end{equation}
The mean of the distribution of this quantity for all the runs
with more than $3\cdot 10^5$ events shows a slight shift ($\approx
0.4\%$) toward negative values. The $rms$ of the distribution is
$\approx 0.009$ or better, depending on target material.

\subsection{The dependence of $\lambda_{mb}$ on the target interaction rate}
\label{lambda_mb_vs_rate}

The dependence of $\lambda_{mb}$, as determined by the method of
Section~\ref{sec:det_of_lambda}, on the target steering hodoscope rate
($R_{hod}$) was checked. In general the dependence should be linear at
sufficiently low interaction rate.

\begin{figure}[h] \centering
\includegraphics*[bb=2 270 520 520,width=\textwidth]{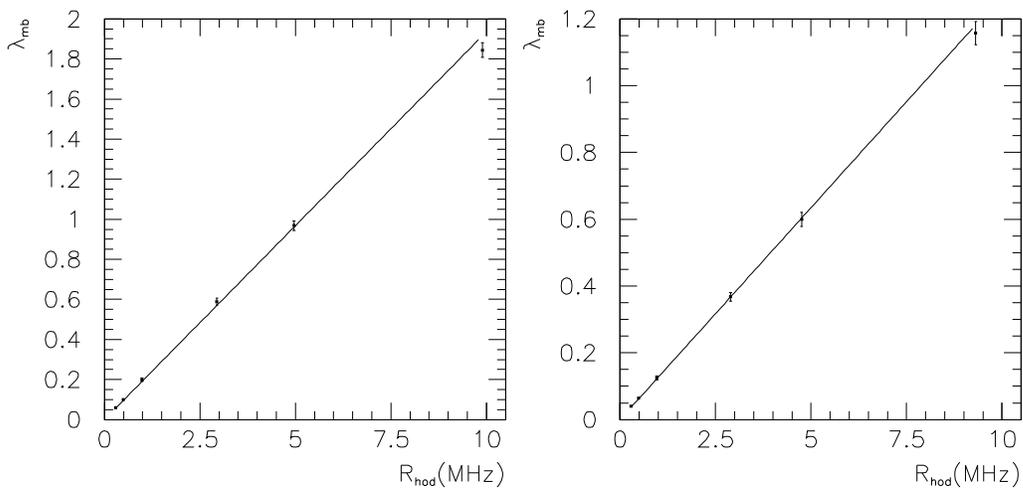}
\caption{\it Test of linear dependence of the $\lambda_{mb}$
values determined with the method described in
Section~\ref{sec:det_of_lambda} with respect to the target
interaction rate measured by the scintillator hodoscope system for
the carbon (left plot) and tungsten (right plot) target
wires.}\label{fig:rette}
\end{figure}

For each of the three target materials (C, Ti and W),
$\lambda_{mb}$ was determined for hodoscope interaction rates of
0.3, 0.5, 1, 3, 5 and 10 MHz. For each target, the resulting
$\lambda_{mb}$ values, excluding the $10\;MHz$ point, were fit to
a straight line.
 The results of the fit for carbon and tungsten target wires
are shown in Fig.~\ref{fig:rette}. The $10 \;MHz$ point lies
below the fit line indicating possible saturation of the
hodoscopes or possibly a breakdown of the Poisson assumption at
high interaction rates. The normalized $\chi^2$ of the linear fit
is about one or better (for all three set of runs) and indicates a
linear relation between interaction rates measured by two very
different techniques for rates up to $5\; MHz$. This in turn
supports the two assumptions made in
Section~\ref{sec:General-remarks-luminos-determi} since the
hodoscope rates do not rely on the Poisson assumption and
compensating non--linearities in the two methods are unlikely.

\input{background.tex}

%% file: background.tex
\subsection{Background estimate}
\label{sub:Backgro-estimat-other-related-systema-uncerta}

As shown in Equation~\ref{eq:fthresh}, the number of recorded events
must be corrected for background events either from fake triggers
(e.g. electronic noise in the RICH) or from coasting beam
interactions (beam-gas interactions and background from interactions
upstream of the target are negligible).  The best method to determine
the fraction of background events is to compare empty and filled
bunches using the random trigger events. For this purpose the
interaction trigger requirements are applied offline to the random
trigger events which are equally distributed over all 220 bunches.
Normalizing to the 180 filled bunches, we obtain

\begin{equation}
\label{eq:fthreshdef}
 f_{bkg}' = \frac{180}{40} \times \left( \frac{N_{tape}(empty\:BX)}{N_{tape}(filled\:BX)} \right)\,.
\end{equation}

A small (approximately 5\%) correction is made to account for
``in-time'' coasting beam interactions which are well synchronized in
time with the detector's integration gates and thus are no different
from ordinary interactions from bunched beam protons.

The resulting values for \mbox{$f_{bkg}$} from the random trigger
sample are summarized in Table~\ref{table:fbkg} and used in
Equation~\ref{eq:luminew}.  To estimate the uncertainty on
\mbox{$f_{bkg}$}, the software trigger thresholds are varied over a
wide range. The uncertainty, $\delta \mathcal{L}_{bkg}$, is obtained
by dividing the difference of the extreme values by $\sqrt{12}$.  The
uncertainties due to the in-time coasting beam correction discussed
above are negligible compared to the uncertainties given in the Table.

\begin{table}[h]
\begin{center}
\begin{tabular}{|r|c|c|}
\hline \hline
      & $f_{bkg}$ & $\left( {\frac{\delta \mathcal{L}}{\mathcal{L}}} \right)_{bkg}$ \\
\hline \hline
    C & 0.031 & 0.018 \\
\hline
   Ti & 0.057 & 0.023 \\
\hline
    W & 0.026 & 0.019 \\
\hline \hline
\end{tabular}
\vspace*{0.3cm}
\caption{\it The fraction of background events \mbox{${f_{bkg}}$} for
each target wire and the relative systematic
uncertainty.
\label{table:fbkg}}
\end{center} 
\end{table}

%% file: dsect.tex
\subsection{Production of $\delta$-rays in the target}
\label{sec:delta}

The presence of $\delta$-rays in the data sample (see
Section~\ref{sec:det_of_lambda}) is both a nuisance, since it
compromises the VDS based methods, and an opportunity for a systematic
check of the luminosity calculation, since the luminosity can be
estimated from the observed rate of $\delta$-ray production. The
results of a study of $\delta$-ray production applied to a run taken
with the carbon target sample are presented and compared to the
luminosity estimates given in Section~\ref{sec:det_of_lambda}.  With
further development, the techniques presented here could be used for a
precise luminosity determination in experiments using thin
targets.

The luminosity for a fixed-target experiment in a proton beam is
proportional to the sum of target path lengths of all protons
($N_{tot}$) which traverse the target:
\begin{equation}
\label{eq:lumfix} \mathcal{L} = \frac{\rho N_A }{A} \cdot
\sum_{i=1}^{N_{tot}} z_i
\end{equation}
where $A$ is the atomic mass of the target material, $N_A$ is
Avogadro's number, $\rho$ is the target density in
($\mathrm{g/cm^3}$) and $z_i$ is the length of the target
traversed by the $i$th proton.

The number of $\delta$-rays ($N_{\delta,prod}$) produced in a kinetic
energy ($T$) interval from $T_{min}$ to $T_{max}$ is proportional
to the same summed target length~\cite{pdg}:
\begin{equation}
\label{eq:ndel}
N_{\delta,prod} = 0.154 \frac{Z}{A} \rho \sum_{i=1}^{N_{tot}} z_i \cdot \int_{T_{min}}^{T_{max}} \frac{dT}{T^2}
        \approx 0.154 \frac{Z}{A} \frac{\rho}{T_{min}} \sum_{i=1}^{N_{tot}} z_i
\end{equation}
where $Z$, $A$, and $\rho$ are the atomic number, atomic mass and
density (in $\mathrm{g/cm^3}$) of the target and $T$ is in MeV.
$T_{max}$ is approximately 475\,GeV for 920\,GeV incident protons.

Combining Equation\,\ref{eq:lumfix} and Equation\,\ref{eq:ndel} results in
the following equation relating luminosity to the number of produced
$\delta$-rays:

\begin{equation}
\label{eq:lumdel} \mathcal{L} = \frac{N_A T_{min}}{0.154 Z} \cdot
N_{\delta,prod} =  \frac{N_A T_{min}}{0.154 Z} \cdot
\frac{N_{\delta,obs}}{\varepsilon_{\delta}},
\end{equation}
where $N_{\delta,obs}$ is the number of observed $\delta$-rays and
$\varepsilon_{\delta}$ is the average probability that the $\delta$-ray
escapes the target and is reconstructed. 

The $\delta$-ray detection efficiency ($\varepsilon_{\delta}$) is
evaluated by Monte Carlo. The $\delta$-rays are generated according to
Equations 27.5 and 27.6 of~\cite{pdg} with a minimum kinetic energy
threshold of 1\,MeV and tracked through the target and detector using
the GEANT3-based~\cite{GEANT} HERA-B simulation program.  
Since the VDS
has acceptance for tracks from the target in the polar angular
interval \mbox{$ 0.01 \lesssim \theta \lesssim 0.7 \,\mathrm{rad}$},
corresponding to a $\delta$-ray momentum range of \mbox{$1.88 \lesssim
  p_\delta \lesssim 10220\,\mathrm{MeV/c}$} (see Equation 27.6 of~\cite{pdg}), the 1\,MeV kinetic energy
threshold corresponds to $\delta$-rays which are well outside the
detector acceptance.  The generated Monte Carlo events are subjected
to the same reconstruction and analysis code (see below) used
for the data.

An average efficiency of $\approx 7\%$ after all cuts is found.  We
estimate a 15\% relative systematic uncertainty on this number coming
from uncertainties in the material distribution in the vertex detector
and from sensitivity to Monte Carlo parameters, in particular to the
minimum kinetic energy cutoff for tracking by GEANT, nominally set to
30\,keV.  (These sources of systematic error could in principle be
greatly reduced by a more precise inventory of detector materials and
by a more thorough study of the tracking of very low momentum
electrons.)

Candidate $\delta$-rays are reconstructed using the standard HERA-B
VDS reconstruction software applied to a pseudo-random-triggered
carbon target run and then searched for in events from filled bunch
crossings which do not pass the $IA$ trigger condition. Distributions of
track parameters and derived quantities from the $\delta$-ray Monte
Carlo and data are found to be in close agreement when segments which
extrapolate to near the average vertex position of hadronic
interactions are removed. Such tracks are typically high momentum
tracks from hadronic $pN$ interactions.

\begin{figure}
\includegraphics*[bb=49 317 756 505,width=\textwidth]{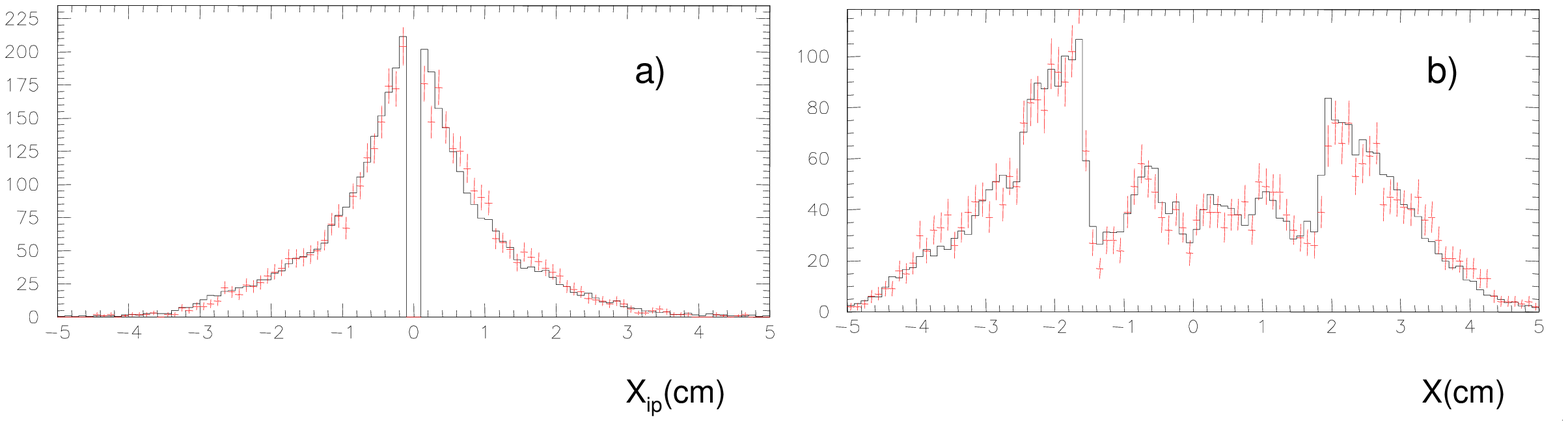}
\caption{\label{delta_fig} \it{a) The distribution of $X_{ip}$ for
selected tracks (see text). b) The distribution of $X$-view impact
points on the first VDS detector layer, for the same tracks used
in part a. In both cases, the data is indicated by the histogram
and the Monte Carlo by the points with error bars.} }
\end{figure}

We define the impact parameter of a track as the difference between
the average position of vertices from hadronic interactions (and
therefore the average impact point of the beam on the wire target) and
the track's position when extrapolated to the $Z$-position of the
target.
Figure~\ref{delta_fig}(a) shows the $X$-view impact
parameter ($X_{ip}$) distribution of reconstructed VDS segments in
non-$IA$ events which contain a single reconstructed segment with
$|X_{ip}| > 1\,\mathrm{mm}$ which originates in the first VDS layer.
The cut on $X_{ip}$ removes a signal from high-momentum tracks from hadronic $pN$
interactions (approximately 1/4 of the removed tracks form a narrow peak above
the relatively broad distribution shown in the figure).  
The data, indicated by the histogram, and the Monte
Carlo, indicated by the points with error bars, are in good agreement.
Note that the width of the $X_{ip}$ distribution is largely determined
by multiple scattering in the VDS and therefore depends on the
momentum of the reconstructed tracks. The close match between data and
Monte Carlo implies that the momentum spectra of reconstructed tracks
in data and the $\delta$-ray Monte Carlo are similar.  Figure~\ref{delta_fig}(b), showing
the distribution of the $X$-view impact points of the same tracks used in
Figure~\ref{delta_fig}(a) at the first VDS layer, also illustrates
the good agreement between data and Monte Carlo. The corresponding
distributions in the $Y$-view also agree well with each other as do 
the distributions of track polar angles. The overall close
agreement strongly suggests that the observed tracks are indeed caused
by $\delta$-rays originating in the target. Further evidence that the
observed tracks are associated with beam protons traversing the target
comes from the greatly reduced rate of such tracks in empty bunch
crossings: $\approx 3\%$ of the rate in filled bunch crossings.

The observed rate of $\delta$-ray candidates is $0.068\pm0.003$ per {\em BX}.
When the same event and track selection criteria are applied to the
minimum bias Monte Carlo, a rate of $5 \cdot 10^{-4}$ candidates per
interaction is observed, or approximately $10^{-4}$ candidates per {\em BX}
for the analyzed run.  The rate of target single diffractive events
(0.004 per {\em BX}, see Table~\ref{table:joaoboris}) is also small compared
to the observed rate. We conclude that the observed tracks cannot be
due to hadronic interactions.

Assuming that the observed tracks are indeed $\delta$-rays,
Equation~\ref{eq:lumdel} gives a luminosity estimate of $633 \pm 28 \pm 95
\;\mathrm{mb^{-1}}$ per {\em BX}, to be compared with the luminosity
estimate of $688 \pm 35\;\mathrm{mb^{-1}}$ computed using the method
described in Section~\ref{sec:det_of_lambda}. The agreement within
errors lends further credence to the hypothesis that the tracks
described in this section are $\delta$-rays from the target and also
serves as a cross-check of the method of
Section~\ref{sec:det_of_lambda}.

%% file: conclusions.tex
As previously noted the uncertainties affecting the total
luminosity measurement are dominated by the systematic
contribution, since each $IA$ trigger run contains enough random
trigger events to make the contribution from statistics
negligible.
 In Table~\ref{tab:doxa} we summarize the
overall relative uncertainty on the total luminosity calculation
\mbox{$\left( {\frac{{\delta{\mathcal{L}_{tot} }
}}{{\mathcal{L}_{tot} }}} \right)$}. In the second column the
uncertainty on $K_A$ is given. This contribution depends on the
present knowledge of the cross sections (see
Table~\ref{table:joaoboris}) and can in principle be improved in
the future. The following three columns list individual
contributions to the systematic uncertainty on detection which are
combined according to Equation~\ref{eq:systdet} to give the total
detection uncertainties shown in the sixth column.

\begin{table}[h]
\begin{center}
\begin{tabular}{|r||c||c|c|c||c||c||c|}
\hline \hline
           &     $\frac{\delta{K_A}}{K_A}$ &   $\frac{\delta{\lambda_{mb}}}{\lambda_{mb}}$ & $\left(\frac{\delta\mathcal{L}}{\mathcal{L}}\right)_{Pois}$ & $\left(\frac{\delta\mathcal{L}}{\mathcal{L}}\right)_{bkg}$  & $\left(\frac{\delta\mathcal{L}_{tot}}{\mathcal{L}_{tot}}\right)_{det}$ & $\frac{\delta\mathcal{L}_{tot}}{\mathcal{L}_{tot}}$ &
$\left(\frac{\delta\mathcal{L}_{tot}}{\mathcal{L}_{tot}}\right)_{uc}$
\\
\hline \hline
   C &    0.023 &   0.039 &   0.009 &   0.018 &   0.044 &    0.050 & 0.039 \\
\hline
  Ti &    0.018 &   0.042 &   0.009 &   0.023 &   0.049 &    0.052 & 0.042 \\
\hline
  W &     0.016 &   0.032 &   0.010 &   0.019 &   0.039 &    0.042 & 0.029 \\
\hline \hline
\end{tabular}
\vspace*{0.3cm}
\caption{\it Values of the contributions to the relative
systematic uncertainty and overall relative systematic uncertainty
on the total luminosity calculation ($ {\frac{{\delta
{\mathcal{L}_{tot} } }}{{\mathcal{L}_{tot} }}} $). The uncorrelated part
$\left(\frac{\delta\mathcal{L}_{tot}}{\mathcal{L}_{tot}}\right)_{uc}$
is given in the last column.
\label{tab:doxa}}
\end{center}
\end{table}

When the method described in this paper is applied to the 2002
HERA-B minimum bias data taking period, the following integrated
luminosities are obtained for each of the three target materials:
\begin{eqnarray}
\mathcal{L}_{tot,C}  = 405.8 \pm  {9.3}   \pm  {17.9}  {\kern 1pt} \;\mu b^{ - 1}  \nonumber\\
\mathcal{L}_{tot,Ti} = 30.9 \pm  {0.6}   \pm  {1.5}  \quad \mu b^{ - 1}  \nonumber\\
\mathcal{L}_{tot,W}  = 38.3 \pm  {0.6}   \pm  {1.5} \quad \mu
b^{-1} \nonumber
\end{eqnarray}
where the first error corresponds to the uncertainty on $K_A$ and
second summarizes the remaining uncertainty mainly due to the HERA-B
experimental conditions. The overall systematic uncertainty can
then be obtained as the quadratic combinations of these two terms.

The method for luminosity measurement described in this paper is
based on the determination of the average number of interactions
per bunch crossing, $\lambda$, and on the knowledge of the total
interaction cross section $\sigma$ (see Equation~\ref{eq:hb_lum}).
The availability of a small fraction (few percent) of events
acquired in parallel to the main stream of data with a completely
unbiased trigger (pseudo-random trigger) was used to evaluate
$\lambda$ on run by run basis.

The measurement of $\lambda$ has been performed by exploiting the
information from a variety of subdetectors, without the use of any
dedicated device. This strategy allowed to perform consistency
checks and to obtain a conservative determination of the
systematic uncertainties of the measurement. 

In case the three data sets will be combined to determine the
A-dependence of a cross section, possible correlations between the
systematic errors have to be taken into account. The correlated error
is dominated by the uncertainty on $\lambda_{mb}$ and the background
correction and is estimated to be $\sim$ 3\%. The correlation coefficients
vary between 0.90 and 0.92 for pairs of wires. 
\footnote{At the time of publication of ~\cite{JPsi,VKstar}, the
correlated error was estimated to be 2\%, rather than the updated and
more accurate value (3\%) presented here.  A recalculation of the values
and errors reported in ~\cite{JPsi,VKstar} using this updated estimate
results in negligable changes.}  
The uncorrelated part of the systematic uncertainty
$\left(\frac{\delta\mathcal{L}_{tot}}{\mathcal{L}_{tot}}\right)_{uc}$
can be found in the last column of Table~\ref{tab:doxa}. 

The same strategy could be applied in future experiments such as
those under construction at the LHC at CERN in Geneva, once the
corresponding cross sections have been measured.

%% file: acknowledgments.tex
We are grateful to the DESY laboratory and to the DESY accelerator
group for their strong support since the conception of the HERA-B
experiment.  The HERA-B experiment would not have been possible
without the enormous effort and commitment of our technical and
administrative staff.